\documentclass[a4paper,preprintnumbers,floatfix,superscriptaddress,pra,twocolumn,showpacs,notitlepage]{revtex4-1}

\usepackage[utf8]{inputenc}
\usepackage[T1]{fontenc}
\usepackage[sc,osf]{mathpazo}
\usepackage{amsmath, amsthm, amssymb,amsfonts}
\usepackage{graphicx}
\usepackage{dcolumn}
\usepackage{bm}
\usepackage{bbm}
\usepackage{hyperref}
\usepackage{comment}
\usepackage{color}

\def\1{\mathbf{1}}
\def\0{\mathbf{0}}

\DeclareMathOperator{\Tr}{Tr}

\newcommand{\ket}[1]{| #1 \rangle}
\newcommand{\bra}[1]{\langle #1 |}

\newcommand{\mean}[1]{\left\langle #1 \right\rangle}

\renewcommand{\rho}{\varrho}

\newcommand{\processnext}[1]{%
  \ifx\listfinish#1\empty\else\listact{#1}\expandafter\processnext\fi}

\newcommand{\figref}[1]{Fig.~\ref{#1}}

{\begin{framed}\begin{small}}
{\end{small}\end{framed}}

\DeclareGraphicsExtensions{.pdf,.png,.jpg}

\makeindex

\begin{document}
\title{Experimental bilocality violation without shared reference frames}
\date{\today}

\author{Francesco Andreoli}
\affiliation{Dipartimento di Fisica - Sapienza Universit\`{a} di Roma, P.le Aldo Moro 5, I-00185 Roma, Italy}

\author{Gonzalo Carvacho}
\affiliation{Dipartimento di Fisica - Sapienza Universit\`{a} di Roma, P.le Aldo Moro 5, I-00185 Roma, Italy}

\author{Luca Santodonato}
\affiliation{Dipartimento di Fisica - Sapienza Universit\`{a} di Roma, P.le Aldo Moro 5, I-00185 Roma, Italy}

\author{Marco Bentivegna}
\affiliation{Dipartimento di Fisica - Sapienza Universit\`{a} di Roma, P.le Aldo Moro 5, I-00185 Roma, Italy}

\author{Rafael Chaves}
\affiliation{International Institute of Physics, Federal University of Rio Grande do Norte, 59070-405 Natal, Brazil}

\author{Fabio Sciarrino}
\email{fabio.sciarrino@uniroma1.it}
\affiliation{Dipartimento di Fisica - Sapienza Universit\`{a} di Roma, P.le Aldo Moro 5, I-00185 Roma, Italy}

\begin{abstract}
Non-classical correlations arising in complex quantum networks are attracting growing interest, both from a fundamental perspective and for potential applications in information processing. In particular, in an entanglement swapping scenario a new kind of correlations arise, the so-called nonbilocal correlations that are incompatible with local realism augmented with the assumption that the sources of states used in the experiment are independent. In practice, however, bilocality tests impose strict constraints on the experimental setup and in particular to presence of shared reference frames between the parties. Here, we experimentally address this point showing that false positive nonbilocal quantum correlations can be observed even though the sources of states are independent. To overcome this problem, we propose and demonstrate a new scheme for the violation of bilocality that does not require shared reference frames and thus constitute an important building block for future investigations of quantum correlations in complex networks.
\end{abstract}

\maketitle

\section{Introduction}Bell's theorem \cite{Bell1964} shows that quantum mechanical correlations can violate the constraints imposed by any classical explanation based on local realism, the phenomenon known as Bell non-locality. To prove the emergence of quantum nonlocal correlations it is sufficient to consider a very simple causal structure, where distant parties perform different measurements on their shares of a quantum entangled system. The classical description of this scenario defines the so-called \textit{local hidden variables} (LHV) models, which have been widely studied \cite{Brunner2014}, up to the recent loophole-free experimental tests \cite{Hensen2015,Shalm2015,Giustina2015} that have conclusively ruled out such causal models as a possible explanation for quantum phenomena. 

Recently, generalizations of Bell's theorem for more complex causal structures have attracted growing attention \cite{Branciard2010,Branciard2012,Fritz2012,Tavakoli2014,Chaves2015PRL,Spekkens2015,Chaves2016PRL,Rosset2016,Tavakoli2016,Ringbauer2016,Wolfe2016inflation}. Exploring the incompatibility of quantum and classical descriptions in more general scenarios is of fundamental importance to understand the role of causality within quantum theory \cite{Leifer2013,Henson2014,cavalcanti2014modifications,Brukner2014,Chaves2015NC,Costa2016,Fritz2016,Horsman2016can} and promises novel routes for quantum information processing in quantum networks of increasing complexity. A new and important feature of such networks, as for example in quantum repeaters \cite{Sangouard2011}, is the fact that typically the correlations between the distant parties are mediated by several independent sources of quantum states. The simplest example is an entanglement swapping experiment \cite{Zukowski1993}, where a central node (Bob) shares two independent quantum states with two other uncorrelated parties (Alice and Charlie) and by jointly measuring the particles in his possession can generate entanglement between them. A precise classical description of this experiment must include the independence of the sources, which introduces the so called bilocality assumption in Bell's theorem \cite{Branciard2010,Branciard2012}. Importantly, as shown in recent experiments \cite{bilocality_nostro,saunders2016experimental}, there are correlations that admit a LHV model but nonetheless are incompatible with the bilocality assumption.   

\begin{figure*}[t]
\includegraphics[width=0.95\linewidth]{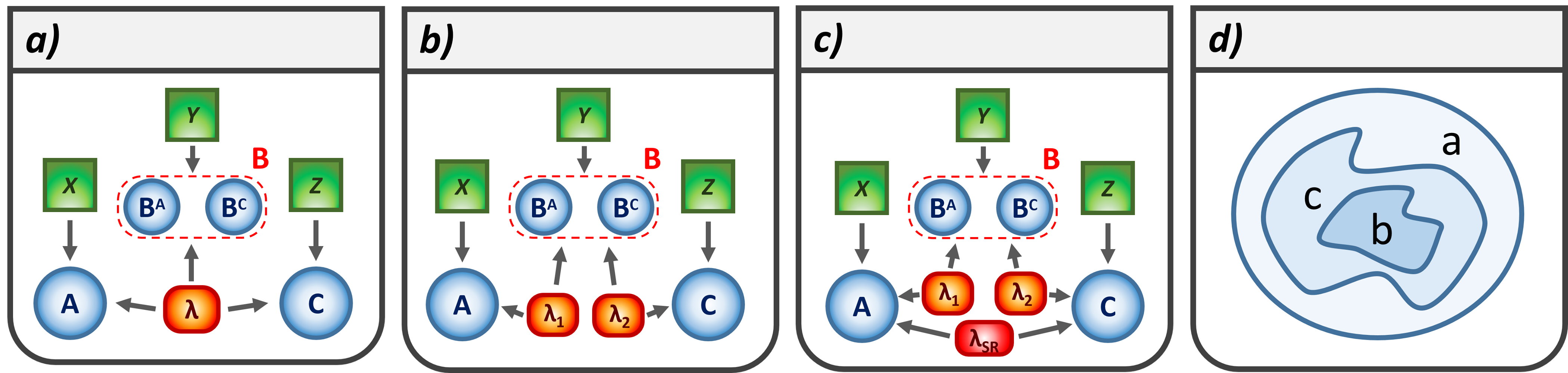}
\caption{\textbf{Directed acyclic graphs (DAGs) \cite{Pearl2009} representing tripartite causal structures.} Nodes represent random variables while arrows represent their causal relations. Blue circles show variables related to measurements outcomes; green squares represent measurement choice inputs and orange blunt rectangles represent hidden variables. \textbf{(a)} Tripartite LHV model where the central node B is made of two substations $B^A$ and $B^C$, driven by the same measurement choice $y$.  \textbf{(b)} BLHV model with the two independent hidden variables $\lambda_1$ and $\lambda_2$ mediating the correlations.  
\textbf{(c)} Classical representation of the previous scenario when allowing for shared randomness between $A$ and $C$ (red rectangle).
\textbf{(d)} Pictorial representation of the structure of the correlations in the three subpanels a, b and c. 
}
\label{fig:DAGS}
\end{figure*}

In this work we focus on related and yet unexplored aspects of the bilocality scenario. First, we experimentally verify a novel route to certify nonbilocal quantum correlations, performing only single qubit operations, as opposed to the results in \cite{bilocality_nostro,saunders2016experimental} that require entangled measurements. We then investigate the importance of the independence between Alice and Charlie, in particular in relation with the necessity of sharing a reference frame (RF) between the parties \cite{Liang2010,Wallman2012,shadbolt2012guaranteed,d2012complete,Brask2013,Wang2016}. While in a usual Bell test shared RFs do not change the causal structure under test, this is no longer true in a bilocality test since they can establish hidden correlations between parties that are assumed to be independent. We experimentally verify this point by showing that nonbilocal correlations may arise even though only classical channels are shared between the parties, thus leading to false positive quantumness. We thus study how this independence can be ensured, by proposing and implementing a scheme which violate a bilocality inequality without the need of shared RFs. Our results thus show that bilocality violation is a feasible and reliable method to certify quantum correlations in a tripartite scenario. 

\section{Witnessing nonbilocality with separable measurements}The classical description of a tripartite Bell scenario depends on the precise causal structure under test. In a usual Bell scenario, one assumes that the correlations between all distant parties are mediated by a single common hidden variable, the standard LHV model, which is depicted in \figref{fig:DAGS}-a.  
However, in a scenario akin to entanglement swapping, the fact that particles are generated in two distinct and separated sources is taken into account by considering two different and causally independent hidden variables $\lambda_1$ and $\lambda_2$, leading to the definition of a \textit{bilocal hidden variable} (BLHV) model (\figref{fig:DAGS}-b). 
If Alice, Bob and Charlie measure two possible dichotomic observables each ($A_0$, $A_1$, $B_0$, $B_1$,  $C_0$ and $C_1$), any correlation compatible with this model respects the inequality \cite{Branciard2012}
\begin{equation}
\label{eq:bilocalineq}
\mathcal{B}=\sqrt{|I|}+\sqrt{|J|} \leq 1,
\end{equation}
with $I=\frac{1}{4} \sum \langle A_{x}B_{0}C_{z} \rangle$, $J=\frac{1}{4}  \sum (-1)^{x+z} \langle A_{x}B_{1}C_{z} \rangle,$ (summing on $x,z=0,1$) and where $\langle A_{x}B_{y}C_{z} \rangle=  \sum (-1)^{a+b+c} p(a,b,c|x,y,z)$ (summing on $a,b,c=0,1$).

Imposing the same causal structure to quantum mechanics we can generate correlations violating inequality \eqref{eq:bilocalineq}, the so-called nonbilocal correlations that are incompatible with a BLHV model. As experimentally shown in Refs. \cite{bilocality_nostro,saunders2016experimental} using photons entangled in polarization, a violation of bilocality can be witnessed even when the experimental data is compatible with a LHV model, thus demonstrating a new form of quantum nonlocality. However, these experiments consider a scenario where the central node B is assumed to perform a complete Bell-state measurement with four possible outcomes, something that is impossible with linear optics \cite{Lutkenhaus1999}. In practice this implies that such realizations have to rely on the precise quantum description of the experiment, that is, we have a device-dependent test of nonbilocality. It is then natural to consider whether a device-independent test of bilocality --where we do not have to rely on the precise description of the measurements but only on the empirical data-- is possible.

Notably it can be shown that QM can exhibit nonbilocal correlations also in case where all parties only perform single qubit operations (i.e. $\sigma_x, \;\sigma_z,\;\sigma_y $ and linear combinations). Such measurements can be performed with linear optics, then allowing for a device-independent test of bilocality. For instance, inequality \eqref{eq:bilocalineq} is violated if (A, B) and (B, C) respectively share a pair of photons in a singlet state and perform the following measurements: $A_0=C_0=(\sigma_x+\sigma_z)/\sqrt{2}$, $A_1=C_1=(\sigma_x-\sigma_z)/\sqrt{2}$ and $B_y=(1-y)(\sigma_{x}\otimes\sigma_x)+y(\sigma_{z}\otimes\sigma_z)$ \cite{Rosset2016, Andreoli2017}. Since all parties only perform single qubit operations this scenario is equivalent to considering outcome B as a function of two individual outcome $B^A$ and $B^C$, both driven by the measurement choice $Y$ (\figref{fig:DAGS}-b).

\begin{figure}[t]
\includegraphics[width=0.96\linewidth]{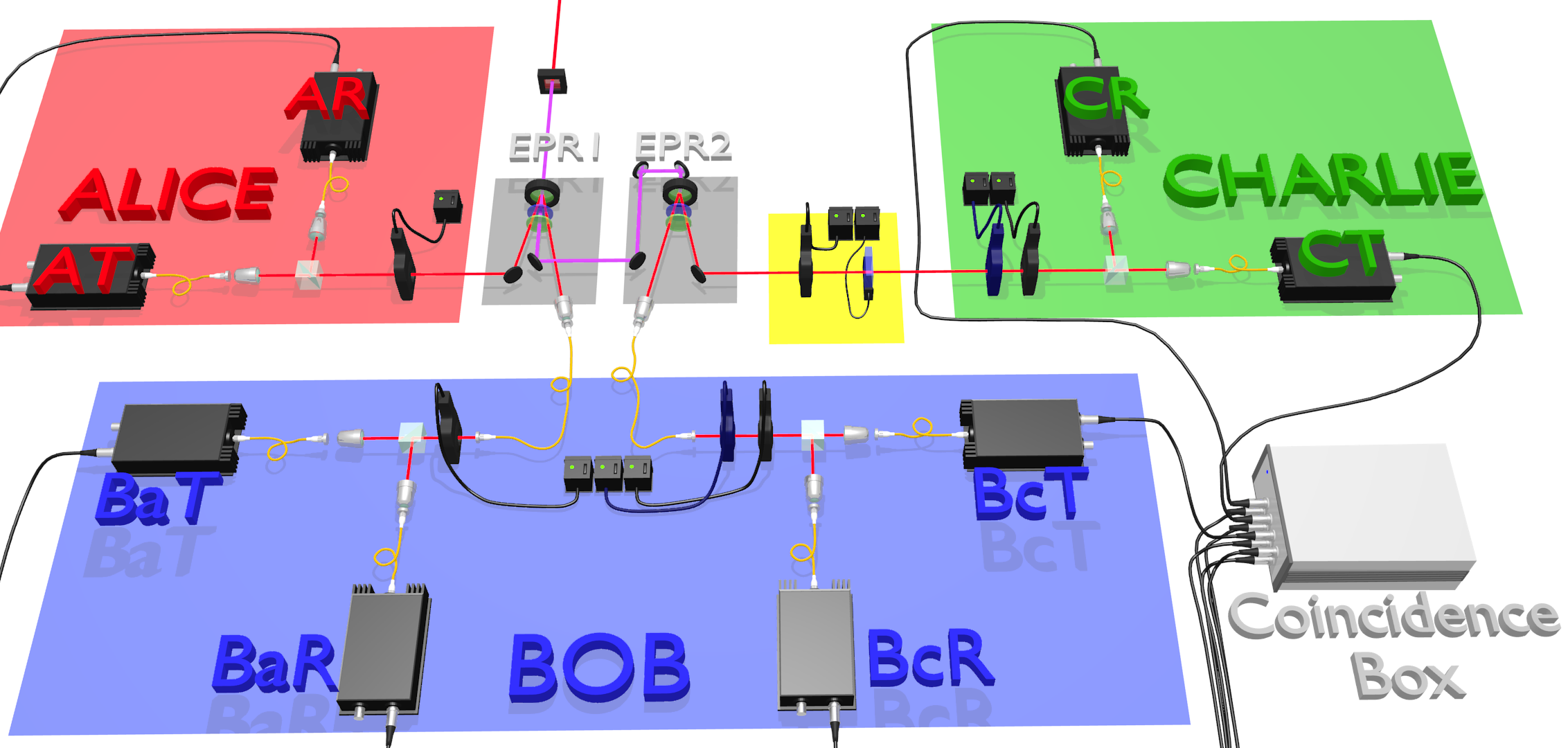}
\caption{\textbf{Experimental setup}. Two separated non-linear crystals (EPR1 and EPR2) generate polarization-entangled photons pairs via spontaneous parametric down conversion (SPDC) process.  
Photons 1 and 4 are directed to Alice's and Charlie's stations respectively, where the analysis in polarization of a particular observable is performed by using a motorized HWP and a Polarizing Beam Splitter (PBS). Photons 2 and 3 are directed to the two different substations $B^A$ and $B^C$ where an analogous analysis is performed.
}
\label{fig:setup}
\end{figure}

We experimentally witness quantum nonbilocality exploiting only separable measurements, using a photonic setup where two independent pairs of photons (1-2 and 3-4 in figure \figref{fig:setup}) are generated by a type-II Spontaneous Parametric Down-Conversion process (SPDC) in two separated crystals. One photon of each source is then directed to one of two different measurement stations (photon 1 to Alice and photon 4 to Charlie) where polarization analysis is performed by using a Half-Wave Plate (HWP) followed by a Polarizing Beam Splitter (PBS). Photons 2 and 3 instead are separately sent to the two substations $B^A$ and $B^C$ where two polarization analysis are performed, analogously to the ones performed in stations A and C. 
Within this scheme, we obtain $\mathcal{B}=1.2712 \pm 0.0042$, witnessing a violation of \eqref{eq:bilocalineq} of almost $65$ sigmas.

\section{Quantum nonbilocality mimicked by shared randomness}When performing a bilocality test, it is essential to ensure that the experiment actually has the causal structure as the one depicted in \figref{fig:DAGS}-b. In particular, one has to guarantee that Alice and Charlie do not have access to any shared randomness. At first, this assumption may seem to be straightforwardly ensured by preparing a quantum state in the form $\rho_{AB}\otimes \rho_{BC}$, indeed guaranteeing that Alice and Charlie are uncorrelated. However, when performing Bell tests, it is usually necessary to establish a shared RF between the measurement devices \cite{Liang2010,Wallman2012,shadbolt2012guaranteed,d2012complete,Brask2013,Wang2016}. While this is generally a practical rather than a fundamental issue, in a bilocality scenario the communication required to establish a RF could also be used to establish undesired correlations between parties that are assumed to be independent. Therefore, in this case, we cannot guarantee that the experimental setup fulfills the bilocality assumption anymore. In fact, as we show next, it is sufficient to allow for some shared randomness between Alice and Charlie (\figref{fig:DAGS}-c) to classically simulate nonbilocal correlations.
 
We experimentally show this phenomenon exploiting two pairs of photons in a state $\rho_{AB}\otimes \rho_{BC}$, where $\rho_{AB}=\rho_{BC}= \sqrt{v}(\ket{HH}\bra{HH}+\ket{VV}\bra{VV})/2+(1-\sqrt{v})\mathbb{I}/4$ (see Appendix A for state generation details). 
Both pairs of photons are in a classical mixture of separable states and thus, if the bilocal causal structure (\figref{fig:DAGS}-b) is imposed, no violation of the inequality \eqref{eq:bilocalineq} should be observed. That is not the case if Alice and Charlie share some underlying correlations, represented in \figref{fig:DAGS}-c by a binary variable $\lambda_{SR}$. Making use of this shared randomness and using the measurement settings $A_0=A_1=C_0=C_1=\sigma_z$ and $B^A_0=B^C_0=B^A_1=B^C_1=\sigma_z$ we can exactly simulate quantum correlations. To that aim, when $\lambda_{SR}=0$ the parties perform a usual bilocality test applying the aforementioned settings. When $\lambda_{SR}=1$, Alice and Charlie apply a post-processing to their outcomes multiplying them by $(-1)^x$ and $(-1)^z$, respectively.
This leads to the two probabilities $p_0\equiv p(a,b,c|x,y,z,\lambda_{SR}=0)$ and $p_1\equiv p(a,b,c|x,y,z,\lambda_{SR}=1)$, where:
\begin{equation}
p_{k}=v\frac{1}{8} \left[ 1+ (-1)^{a+c+b+k(x+z)} \right] +(1-v)\frac{1}{8}
\label{Prob_i}
\end{equation}
These two distributions are bilocal for all $v$. However, using their shared randomness, Alice and Charlie can mix their strategies in a correlated manner and obtain $p(a,b,c|x,y,z)=(p_{0}+p_{1})/2$, which violate \eqref{eq:bilocalineq} for any $v > 1/2 $.

\begin{figure}[t]
\includegraphics[width=0.96\linewidth]{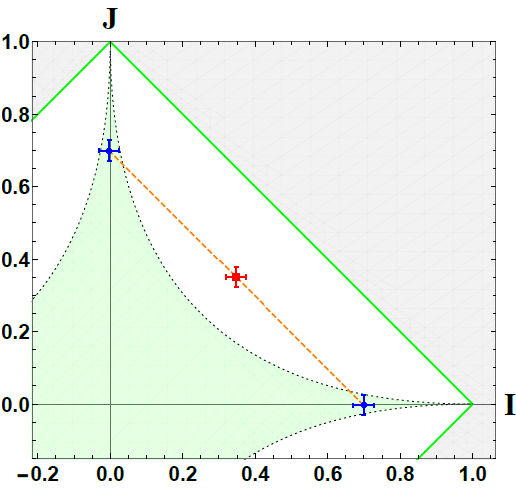}
\caption{\textbf{Nonbilocal correlations from shared randomness.} Green shaded area within the dotted lines shows the bilocal set in the $I, \;J$ plane, while the bold green line bounds the local set (that includes the bilocal one). Blue circle points show the experimental results for a fixed distribution \eqref{Prob_i}, while the red square point represents the equal mixing of these two points (i.e. shared randomness). Error bars are due to poissonian statistics.}
\label{fig:bil}
\end{figure}

Our experimental results are shown in \figref{fig:bil}. The blue circles represent the case of no shared randomness, with the measurement outputs defined by the $\lambda_0$ or $\lambda_1$ prescriptions. As expected, these points are inside the bilocal set, showing an experimental value of $\mathcal{B}=0.892 \pm 0.062$, which is compatible with \eqref{eq:bilocalineq} by almost 2 sigma. The red square shows how the mixing of these two probabilities gives a nonbilocal point with $\mathcal{B}=1.180\pm 0.034$, meaning a bilocality violation of more than $5$ sigma. This shows that the independent preparation of two quantum states $\rho_{AB}$ and $\rho_{BC}$ is not sufficient to justify the bilocality assumption and thus guarantee the observation of truly quantum correlations in this tripartite network.

\section{Violation of bilocality without shared reference frames between some of the stations} In view of this result, any experimental observation of nonbilocality requiring the use of shared RFs (that may generate correlations between Alice and Charlie) is highly unsatisfactory. Thus, to turn the violation of a bilocality inequality into a reliable test of non-classical behaviour is crucial to ask:
is it possible to observe nonbilocal correlations even in the absence of shared RFs? To answer in positive to this question, we consider a few relevant situations. In the following we focus on a scenario where only Alice and Bob share a RF. We describe a new scheme allowing for the violation of bilocality in this case as well as an experimental implementation of it. In the next section we also propose a scheme allowing for witnessing nonbilocality in the absence of any shared RF and in the Appendix we finally address the case when no local calibration of measurement devices is ensured.

First, we notice that defining $\vec{a}_i,\;\vec{b}^A_i,\;\vec{b}^C_i,\;\vec{c}_i$ as the measurements vectors associated to $A_i,\;B^A_i,\;B^C_i$ and $C_i$, the bilocality parameter $\mathcal{B}$ can be written as
\begin{equation}
\label{eq:B_general_measurement_form}
\begin{array}{c}
\mathcal{B}=\dfrac{1}{2}\sqrt{\big| (\vec{a}_0+\vec{a}_1)\cdot T_{\rho_{AB}}\vec{b}^A_0 \big| \;\;\big| \vec{b}^C_0\cdot T_{\rho_{BC}}(\vec{c}_0+\vec{c}_1) \big|}+\\\\
\dfrac{1}{2}\sqrt{\big| (\vec{a}_0-\vec{a}_1)\cdot T_{\rho_{AB}}\vec{b}^A_1 \big| \;\;\big| \vec{b}^C_1\cdot T_{\rho_{BC}}(\vec{c}_0-\vec{c}_1) \big|},
\end{array}
\end{equation}
where the matrix $T_{\rho_{AB}}$ has components $(t_{\rho_{AB}})_{ij}=Tr \big[ \rho_{AB}\; \sigma_i \otimes \sigma_j \big] $, and similarly for $T_{\rho_{BC}}$. Equation \eqref{eq:B_general_measurement_form} makes clear that the value of $\mathcal{B}$ is simply given by the Bloch sphere geometry between the vectors $\vec{a}_i$ and $\vec{b}^A_j$ and between $\vec{c}_i$ and $\vec{b}^C_j$. This implies that even if $A$ and $B^A$ share a reference frame which is different from the one shared between $C$ and $B^C$ (that is, even though stations $A$ and $C$ do no share a common RF) we can still obtain maximum violation of \eqref{eq:bilocalineq}. This scenario can be of particular experimental interest when two different and independent channels are used to implement the calibration between the $A$ and $B^A$ and $C$ and $B^C$ measurement devices. However, it still depends on the assumption that no internal calibration between $B^A$ and $B^C$ is taking place.

\begin{figure*}[t]
  \includegraphics[width=0.98\textwidth]{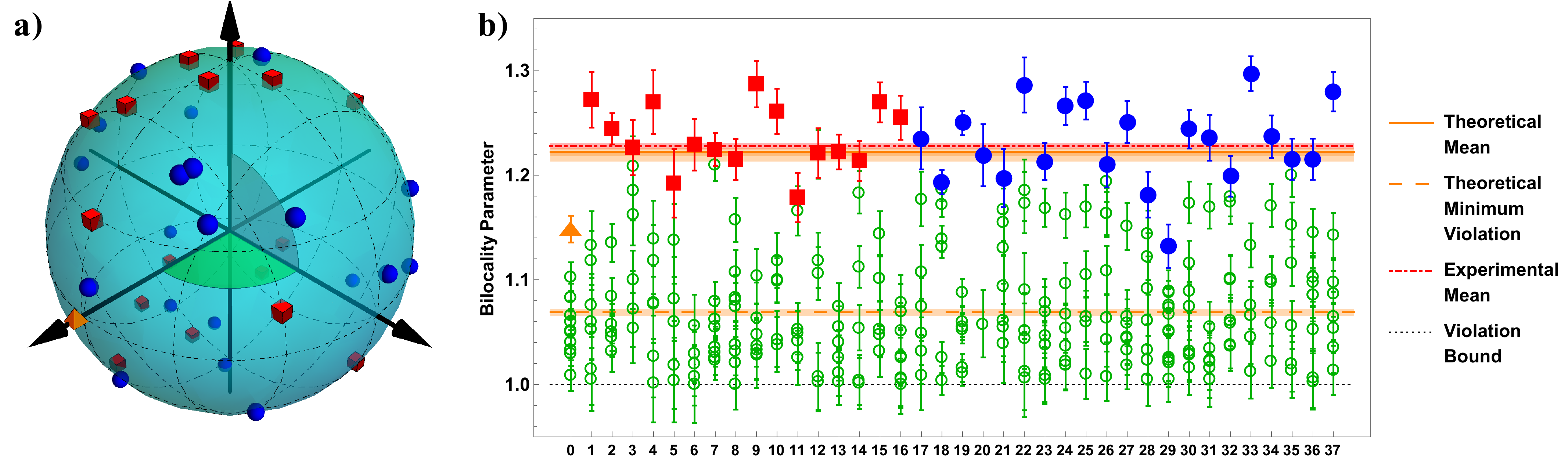}
  \caption{\textbf{Experimental test of bilocality with random reference frames between stations $B^C$ and $C$.}
\textbf{a)} Distribution on the Bloch sphere of the performed unitaries for all the experimental points. Red cubes represent points chosen with a uniform pseudo-random distribution, while blue spheres were chosen using the human-random strings provided by the BIG Bell Test \cite{BIG_Bell_Test}. The orange pyramid depicts the identity unitary. \textbf{b)} Amount of violation for all the different randomized unitaries. Given a unitary transformation $U_i$ green empty circles show the parameter $\mathcal{B}$ for those inequalities (out of 81, considering the different settings and symmetries of \eqref{eq:bilocalineq}) which violate the bilocality condition, except for the maximal violation which is represented with a orange triangle, red squares or blue disks (corresponding, respectively, to the pyramid, cubes and spheres of subpanel a). Error bars indicates one standard deviation of uncertainty, due to Poissonian statistics. The orange straight (dashed) line shows the theoretical mean (minimum) violation expected for the maximum points (orange triangle, red squares and blue disks) considering our subset of unitaries, while the red dot-dashed line represents their experimental mean value. The shaded areas display one standard deviation of uncertainty, both for the experimental mean value and for the theoretical expectations due to the experimental estimation of noise. The black dotted line indicates the bilocality violation bound.
}
\label{fig:Main_experimental_plot}
\end{figure*}

A more satisfactory situation from the device-independence perspective is one where only $A$ and $B^A$ share common a reference frame, thus guaranteeing that no correlation will be established between $A$ and $C$. To tackle this case let us consider $\rho_{AB}=\rho_{BC}=\ket{\psi^-}\bra{\psi^-}$ and $A_x=(\sigma_z+(-1)^x \sigma_x)/\sqrt{2},\;B^A_y=y\sigma_x+(1-y)\sigma_z$. From equation \eqref{eq:B_general_measurement_form} one can show that if $C$ and $B^C$ perform 3 different measurements belonging to an orthogonal triad ($\{\vec{c}_j\}$ or $\{\vec{b}_k\}$), then the following statement holds
\begin{equation}
\label{eq:NO_frame_statement}
\begin{array}{c}
\forall\;\text{R.F.s between $C$ and $B^C$,}\;\;\;\exists \vec{c}_{\gamma}\;\vec{c}_{\gamma'}\in \{\vec{c}_j\} \\\\ \& \;\;\vec{b}^C_{\beta_C}\;\vec{b}^C_{\beta'_C}\in \{\vec{b}^C_k\} \;\;\text{such that}	\\\\
\;\mathcal{B}(\vec{c}_{\gamma},\vec{c}_{\gamma'},\vec{b}^C_{\beta_C},\vec{b}^C_{\beta'_C})\geq\sqrt[4]{2}\sim 1.19,
\end{array}
\end{equation}
where $\{\vec{c}_j\}$ and $\{\vec{b}^C_k\}$ define orthonormal triads.
\begin{proof} If $\rho_{AB}=\rho_{BC}=\ket{\psi^-}\bra{\psi^-}$ and $A_x=(\sigma_z+(-1)^x \sigma_x)/\sqrt{2},\;B^A_y=y\sigma_x+(1-y)\sigma_z$, from equation (3) of the main text we get
\begin{equation}
\label{eq:B_general_measurement_form_A_fixed}
\begin{array}{c}
\mathcal{B}=\dfrac{\sqrt[4]{2}}{2}\big(\sqrt{\big|-\vec{b}^C_0\cdot(\vec{c}_0+\vec{c}_1)\big| }+
\sqrt{ \big|-\vec{b}^C_1\cdot (\vec{c}_0-\vec{c}_1) \big|}\big)= \\\\

\dfrac{\sqrt[4]{2}}{2}\big(\sqrt{\big|E_{00}+E_{10}\big| }+
\sqrt{ \big|E_{11}-E_{01} \big|}\big),\end{array}
\end{equation}
where we introduced the correlator $E_{jk}=-\vec{c}_j\cdot\vec{b}^C_k$. Let $C$ and $B^C$ perform 3 different measurements belonging to an orthogonal triad ($\{\vec{c}_j\}$ or $\{\vec{b}_k\}$). It was shown \cite{shadbolt2012guaranteed} that in this scenario two couples of measurements exist such that the following inequalities hold
\begin{equation}
\begin{array}{c}
\label{eq:ineq_3meas_1}
%E_{00}E_{11}-E_{01}E_{10}\geq E_{00},\\\\
%E_{01}E_{10}\leq 0,\\\\
E_{00}+E_{10}\geq 1,\\\\
E_{11}-E_{01}\geq 1,
\end{array}
\end{equation}
%and that we can choose $E_{01}\leq 0$ and $E_{10}\geq 0$ without loss of generality (it is easy to figure it out since experimentally it is the same to perform $\vec{c}_{1}$ or $-\vec{c}_{1}$ and, thus, it is always possible to exchange $E_{10}$ and $E_{01}$ in equation \ref{eq:B_general_measurement_form_A_fixed}). 
leading to
\begin{equation}
\begin{array}{c}
\label{eq:B2233_minimum_bound}
\mathcal{B}=\dfrac{\sqrt[4]{2}}{2}\big(\sqrt{\big|E_{00}+E_{10}\big| }+
\sqrt{ \big|E_{11}-E_{01} \big|}\big)\geq
%\\\\\dfrac{\sqrt[4]{2}}{2}\big(1+\sqrt{ \big| 1 -2 E_{01} \big|}\big)\geq 
\sqrt[4]{2}.
\end{array}
\end{equation}
\end{proof}
This statement shows thus that is always possible to observe a bilocality violation regardless of the RFs chosen in stations $C$ and $B^C$, with a minimum value given by $\mathcal{B}_{min}\geq\sqrt[4]{2}\sim 1.19$.

We experimentally addressed this point by using a HWP followed by a Liquid Crystal on the path leading to station $C$. These two elements work as a black-box which allows us to rotate one of the two qubits all over its Bloch sphere, before it reaches the measurement station. By choosing random settings, it is thus possible to explore a fair subset of all the possible random unitaries which may occur when stations $C$ and $B^C$ do not share a common reference frame (see Appendix B for further details).

Our experimental results are shown in figure \ref{fig:Main_experimental_plot}. Red cubes/squares represent random settings chosen accordingly to a uniform distribution over the Bloch sphere, while blue spheres/disks exploit human randomness provided by the BIG Bell Test experiment \cite{BIG_Bell_Test}. Figure \ref{fig:Main_experimental_plot}-a shows these shifts on the Bloch sphere, while figure \ref{fig:Main_experimental_plot}-b displays, with red squares and blue disks, the best $\mathcal{B}$ value obtained considering all settings combinations (and symmetries of inequality \eqref{eq:bilocalineq}), for each random unitaries. The red dot-dashed line shows the mean value of all these points, which is in good agreement with the theoretical expected value (considering our amount of noise) for our subset of unitaries (orange line). We found at least one value higher than the expected lower bound of $\sqrt[4]{2}$ (corrected with our noise) for all random unitaries, which is represented by the orange dashed line. It is remarkable that, at the cost of performing $3$ instead of $2$ measurements in $C$ and $B^C$, we can still obtain bilocality violations even when one of the parties shares no RF at all with the other parties.

\section{Complete absence of shared references frames}We will analyze now the case when none of the parties share a reference frame with each other. We will start proving the following statement 
\begin{equation}
\label{eq:NO_frame_statement_ABC}
\begin{array}{c}
\forall\;\text{misaligned R.F.s}\;\;\;\exists \vec{a}_{\alpha}\;\vec{a}_{\alpha'}\in \{\vec{a}_i\} \;\;\& \;\;\vec{b}^A_{\beta_A}\;\vec{b}^A_{\beta'_A}\in \{\vec{b}^A_n\} \\\\\ \&\;\;\vec{c}_{\gamma}\;\vec{c}_{\gamma'}\in \{\vec{c}_j\} \;\;\& \;\;\vec{b}^C_{\beta_C}\;\vec{b}^C_{\beta'_C}\in \{\vec{b}^C_k\} \;\;\text{such that}\\\\
\;\mathcal{B}(\vec{a}_{\alpha},\vec{a}_{\alpha'},\vec{b}^A_{\beta_A},\vec{b}^A_{\beta_A'},\vec{c}_{\gamma},\vec{c}_{\gamma'},\vec{b}^C_{\beta_C},\vec{b}^C_{\beta'_C})>1.
\end{array}
\end{equation}

\begin{proof} 
Let us define $E^A_{jk}=-\vec{a}_j\cdot \vec{b}^A_k$ and $E^C_{jk}=-\vec{c}_j\cdot \vec{b}^C_k$. The bilocal parameter $\mathcal{B}$ can be expressed as

\begin{equation}
\begin{array}{c}
\mathcal{B}=\dfrac{1}{2}\big(\sqrt{\big| E^A_{00}+E^A_{10}\big| }\sqrt{\big| E^C_{00}+E^C_{10}\big| }+\\\\
\sqrt{ \big| E^A_{11}-E^A_{01} \big|}\sqrt{ \big| E^C_{11}-E^C_{01} \big|}\big).
\end{array}
\end{equation}

We can make use of inequalities \ref{eq:ineq_3meas_1}, which are valid for both $E^A_{jk}$ and $E^C_{jk}$, in order to show that

\begin{equation}
\label{eq:No_ref_at_all_ineq}
\begin{array}{c}
\mathcal{B}^2=  \dfrac{1}{4} \big(\; \big| E^A_{00}+E^A_{10}\big| \big| E^C_{00}+E^C_{10}\big| + \big| E^A_{11}-E^A_{01} \big|\big| E^C_{11}-E^C_{01} \big| +\\\\
2\sqrt{\big| E^A_{00}+E^A_{10}\big|\big| E^A_{11}-E^A_{01} \big|\big| E^C_{00}+E^C_{10}\big|\big| E^C_{11}-E^C_{01} \big|}\;\big) \geq\\\\
\dfrac{1}{4} \big( \big| E^A_{00}+E^A_{10}\big| + \big| E^A_{11}-E^A_{01} \big| +2\big)>1,
\end{array}
\end{equation}
where, in the last step, we used the fact that $E^A_{00}+E^A_{10}+ E^A_{11}-E^A_{01}>2$ in case $\{\vec{a}_i\}$ and $\{\vec{b}^A_n\}$ are not aligned reference frames \cite{shadbolt2012guaranteed}. In case $\{\vec{a}_i\}$ and $\{\vec{b}^A_n\}$ are aligned then it is possible to perform analogous calculations which still prove the theorem under the assumption that $\{\vec{c}_j\}$ and $\{\vec{b}^C_k\}$ are not aligned. We stress that, for random triads, the probability of perfect alignment between all the triads is null.
\end{proof}

This statement proves that it is still possible to witness a bilocality violation, even if none of the parties shares a common reference frame, at the only cost of performing 3 measurements per station belonging to an orthogonal triad. Numerical analysis of this case was also performed, and the results are in agreement with the analytical conclusions. All these calculations, however, rely on the presence of local calibration in all of the stations, which allows to perform orthogonal measurements. Without this assumption one can only have a certain probability of witnessing a bilocality violation, which can be increased performing more measurements per party (see Appendix B). \\

\section{Summary and Conclusions}Generalizations of Bell's theorem for complex networks open novel routes both from fundamental and applied perspectives. Understanding the role of causality within quantum mechanics remains a notoriously thorny issue to which causal structures beyond the one originally devised by Bell can offer new insights \cite{Henson2014,Chaves2015NC,Costa2016,Fritz2016}. Further, these more complex networks also have the potential to increase our capabilities to process information in a non-classical way, since they allow for the emergence of new kinds of quantum nonlocal correlations, in particular the so-called nonbilocality. From the experimental side, these new scenarios introduce new possibilities but also new challenges to be circumvented.

In this work we have taken the next step in the experimental investigation of the bilocality scenario. First, we have shown that one can violate a bilocality inequality using only separable measurements, as opposed to recent experiments \cite{bilocality_nostro,saunders2016experimental} that require full Bell-state measurements and can only be performed using linear optics if further assumptions are employed, being thus unsatisfactory from a device-independent perspective. Second, we experimentally demonstrated how the presence of shared randomness between Alice and Charlie can be used to mimic quantum nonbilocal correlations with classical mixtures of separable states. This leads to false positives on the violation of bilocality and impose strict control on the use of shared reference frames between the distant parties involved in the Bell test. Third, in order to deal with this last point, we developed and experimentally verified novel procedures for witnessing nonbilocality even in the absence of common reference frames. Altogether, we believe these results constitute a building block for the future analysis of the non-classical correlations arising in complex quantum networks.

\acknowledgments

G.C. thanks  Becas  Chile  and  Conicyt for  a  doctoral  fellowship. This work was supported by the ERC-Starting Grant 3D-QUEST (3D-Quantum Integrated Optical Simulation; grant agreement no. 307783): http://www.3dquest.eu, the Brazilian ministries MEC and MCTIC and the FQXi Fund.

%\section{Appendix} In this appendix we report the detailed information related with the generation of separable states, an insight on the case when only A and $B^A$ share a reference frame and we also provide a noise parameter which takes into account all the noise that can be present in our experimental realization.

\appendix

\begin{figure*}[t]
\includegraphics[width=0.95\linewidth]{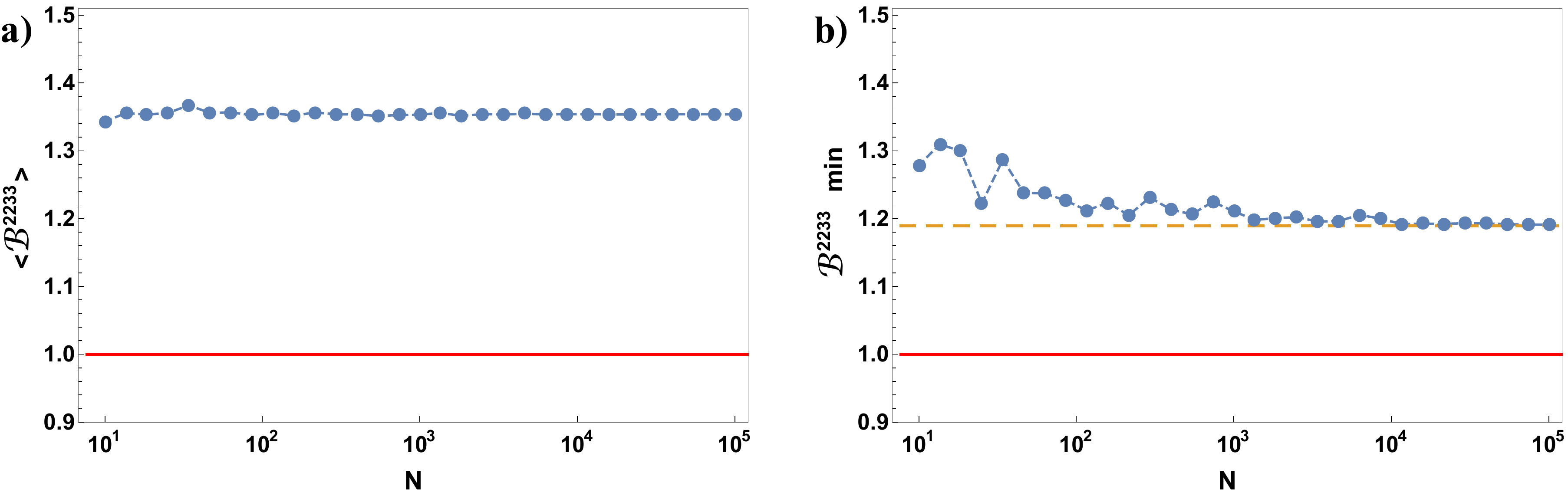}
\caption{\textbf{Numerical analysis of the case when only $A$ and $B^A$ have a common reference frame.}  \textbf{a-b) }Mean (minimum) violation obtained for a growing population of $N$ random misaligned frames. Each point refers to a population which is totally independent from the other ones. The red line shows the violation bound ($1.0$), while the orange dashed line shows the minimum bound of $\sqrt[4]{2}$ derived in \eqref{eq:B2233_minimum_bound}.}
\label{fig_Exp2_Theo-Case2233-3Parameters}
\end{figure*}

\begin{figure*}[t]
\includegraphics[width=0.87\linewidth]{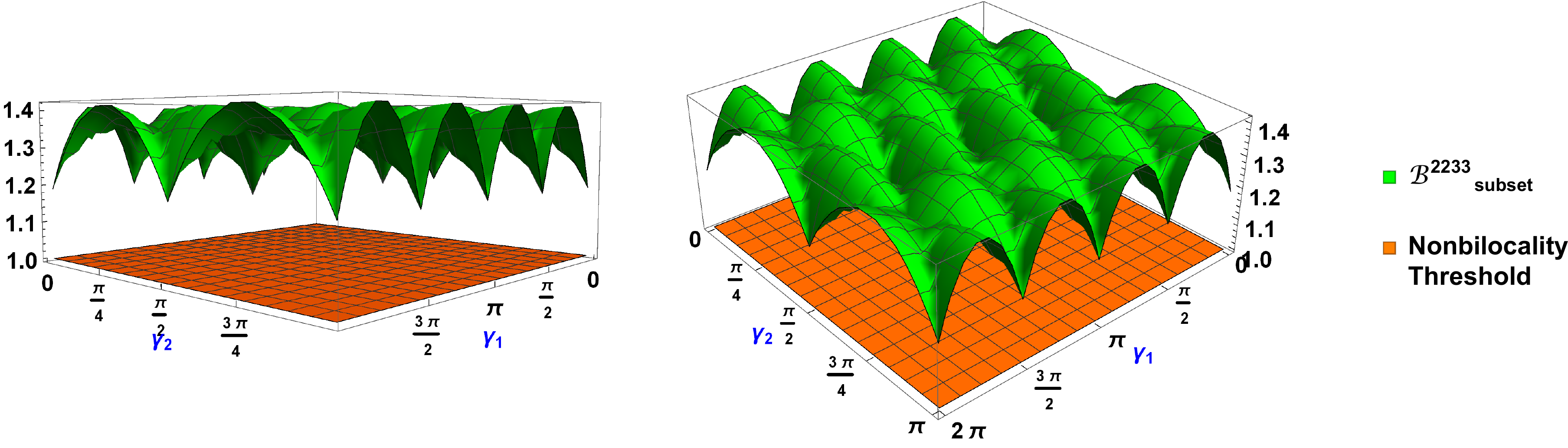}
\caption{\textbf{Numerical analysis of the case $\mathcal{B}^{2233}$ with our experimental subset of random unitaries.} The green surface display $\mathcal{B}^{2233}_{subset}$ for all the possible values of $\gamma_1$ and $\gamma_2$, while the orange plane shows the nonbilocality threshold.}
\label{fig:ABrefOK_BCrefNO}
\end{figure*}

\section{Generating separable states}
%\textit{Generating two independent separable states} - 
We generate the two states $\rho_{AB}=\rho_{BC}=(\ket{HH}\bra{HH}+\ket{VV}\bra{VV})/2$ setting the compensation HWPs placed after the BBO crystals to $0^{\circ}$. Indeed, in this way, a decoherence effect can be exploited in order to destroy the entanglement between the photons. This walk-off effect separates the $H$ and $V$ polarizations in two different paths, which can be taken into account by considering an additional path degree of freedom $\zeta$ for the photons, which isn't experimentally witnessed. In the case where the walk-off is compensated, the states $\rho_{AB}$ and $\rho_{CB}$ will be
\begin{eqnarray}
\rho_{XB}=\Tr_{\zeta}[\;\ket{\psi^-}\ket{\zeta_1\;\zeta_1}\bra{\psi^-}\bra{\zeta_1\;\zeta_1}\;]=\ket{\psi^-}\bra{\psi^-},
\end{eqnarray} 
where $X$ may represent either $A$ or $C$. On the other hand, setting the compensation HWP of photons 1 and 4 to $0^{\circ}$, we have
\begin{eqnarray}
&&\ket{\eta}\equiv\dfrac{1}{\sqrt{2}}(\ket{HH}\ket{\zeta_1\;\zeta_1}-\ket{VV}\ket{\zeta_2\;\zeta_1}),\\
\nonumber 
&& \rho_{XB}=\Tr_{\zeta}[\;\ket{\eta}\bra{\eta}\;]=\dfrac{1}{2}(\ket{HH}\bra{HH}+\ket{VV}\bra{VV}).
\end{eqnarray} 
The initial quantum state is thus in the form $\rho_{ABC}=\rho_{AB}\otimes \rho_{BC}$ where 
\begin{equation}
\label{eq:separable_state}\rho_{AB}=\rho_{BC}= \dfrac{\sqrt{v}}{2}(\ket{HH}\bra{HH}+\ket{VV}\bra{VV})+(1-\sqrt{v})\dfrac{\mathbb{I}}{4},
\end{equation}
which takes into account the presence of white noise.

\begin{figure*}[t]
\includegraphics[width=0.96\linewidth]{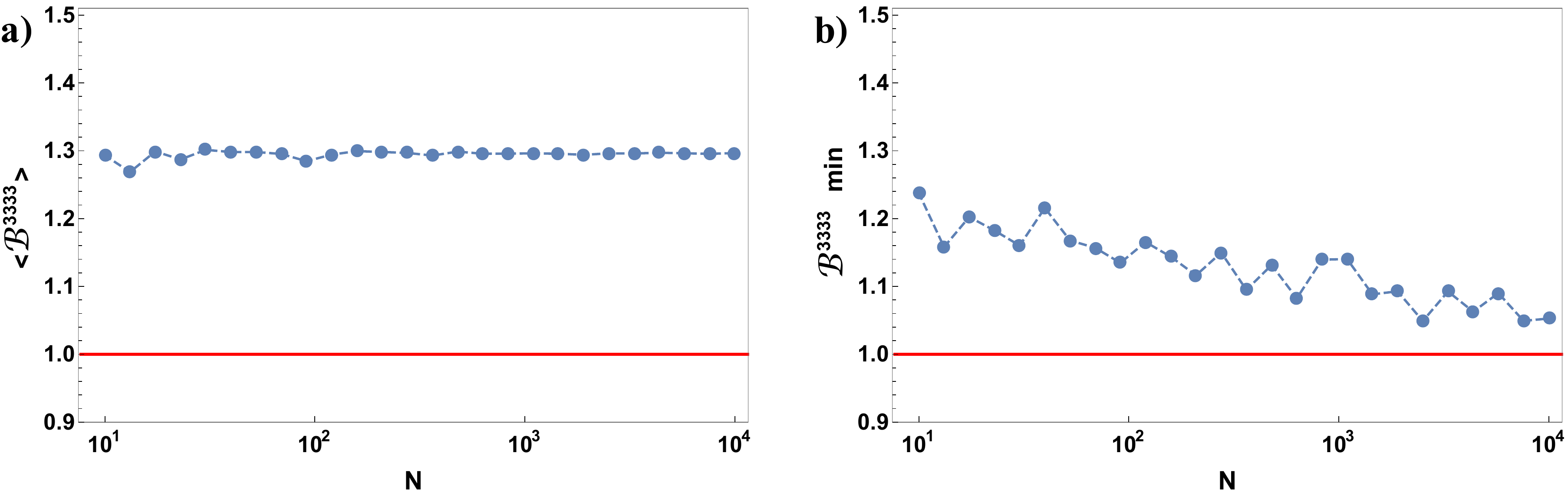}
\caption{\textbf{Numerical analysis of the case when none of the parties have a common reference frame.}  \textbf{a-b) }Mean (minimum) violation obtained for a growing population of $N$ random misaligned frames. Each point is refers to a population which is totally independent from the other ones. The red line shows the violation bound (1.0).}
\label{fig:NoRefAtAll}
\end{figure*}

\section{Noise estimation}
We defined the noise parameter $0 \leq V\leq 1$, by assuming that the noise which can be present in our experimental realization acts as a generic white noise (i.e. isotropic depolarization). Indeed if one considers the two Werner states
\begin{equation}
\begin{array}{c}
\rho_{AB}=v_{AB}\ket{\psi^-}\bra{\psi^-}+(1-v_{AB})\dfrac{\mathbb{I}}{4},\\\\
\rho_{BC}=v_{BC}\ket{\psi^-}\bra{\psi^-}+(1-v_{BC})\dfrac{\mathbb{I}}{4},
\end{array}
\end{equation}
instead of two singlet states, then the expected bilocality parameter becomes $\mathcal{B}_{exp}=V \mathcal{B}_{theo}$, where $V=\sqrt{v_{AB}v_{BC}}$. 

We then took into account the experimental bilocality violation that we obtained (paragraph \textit{Witnessing nonbilocality with separable measurements} of the main text) in the case where a common reference frame for all parties was present. This led to the estimation

\begin{equation}
V=\dfrac{\mathcal{B}_{exp}}{\mathcal{B}_{theo}}=\dfrac{(1.2712\pm 0.0042)}{\sqrt{2}}=0.8989\pm 0.0030.
\end{equation}

\section{Numerical simulations}
\textit{Shared reference frame between Alice and Bob} - We will provide a further insight on the first case discussed in the main text, the one when only $A$ and $B^A$ share a reference frame. Let us call $\mathcal{B}^{2233}$ the bilocality parameter for the case where $A$ and $B^A$ perform 2 measurements, while $C$ and $B^C$ can perform 3 measurements belonging to an orthogonal triad.
%We performed numerical analysis exploiting the fact that only the relative shift between $C$'s and $B^C$'s reference frames is decisive to the value of $\mathcal{B}^{2233}$ (where we denoted with ). It is thus possible to fix one of the two reference frames and consider a generic rotation of the other one. 
In order to perform numerical simulations, one has to consider that a generic rotation in a 3-dimensional space needs three parameters to be identified (two for the choice of a rotation axis and one for the angle of rotation). Equivalently we can consider, without loss of generality \cite{Wallman2012a}, the two frames
\begin{equation}
\begin{array}{l}
x^{BC}=(\sin \beta^C, \; -\cos \beta^C, \; 0),\\\\
y^{BC}=(\cos \beta^C, \; \sin \beta^C, \; 0),\\\\
z^{BC}=(0,\;0,\;1),\\\\
x^{C}=(\sin \gamma_1 \cos \gamma_2 ,\; -\cos \gamma_1 ,\; -\sin\gamma_1 \sin \gamma_2),\\\\
y^{C}=(\cos \gamma_1 \cos \gamma_2 ,\; \sin \gamma_1 ,\; -\cos\gamma_1 \sin \gamma_2),\\\\
z^{C}=(\sin\gamma_2 ,\; 0, \;\cos\gamma_2),
\end{array}
\end{equation}
We can then analyze all the combinations of 2 measurements for $C$ and 2 for $B^C$ and, for each random value of $\beta^C$, $\gamma_1$ and $\gamma_2$, we can take the one which maximizes $\mathcal{B}^{2233}$. We performed several independent runs, with a growing population $N$ of random settings, leading to \figref{fig_Exp2_Theo-Case2233-3Parameters}. For a population of $10^5$ random cases we obtained a mean value (for the maximum violation, considering all combination of settings and the symmetries of the bilocality inequality) of $ \mean{\mathcal{B}^{2233}}\sim 1.345$, while from \figref{fig_Exp2_Theo-Case2233-3Parameters}-b it is evident that the lower bound is given by $\sqrt[4]{2}$.

In our experimental setup we are able only to explore a subset of the random rotations which may occur without the presence of a shared reference frame between $C$ and $B^C$. We indeed are able to randomly rotate one of the two entangled qubit all over its Bloch sphere, leading to the two triads
\begin{equation}
\nonumber
\begin{array}{l}
x^{BC}_{subset}=(1, \;0, \; 0),\\\\
y^{BC}_{subset}=(0, \; 1, \; 0),\\\\
z^{BC}_{subset}=(0,\;0,\;1),\\\\
x^{C}_{subset}=(\sin \gamma_1 \cos \gamma_2 ,\; -cos \gamma_1 ,\; -\sin\gamma_1 sin \gamma_2),\\\\
y^{C}_{subset}=(\cos \gamma_1 \cos \gamma_2 ,\; \sin \gamma_1 ,\; -\cos\gamma_1 \sin \gamma_2),\\\\
z^{C}_{subset}=(\sin\gamma_2 ,\; 0, \;\cos\gamma_2).
\end{array}
\end{equation}
This subset of random unitaries depends only on two parameters ($\gamma_1$ and $\gamma_2$) and can be represented by a surface in a 3 dimensional space (\figref{fig:ABrefOK_BCrefNO}). We evaluated the mean violation for this case by numerically integrating this surface, obtaining $\mean{\mathcal{B}^{2233}_{subset}}\sim 1.359$ and a lower violation of $\sqrt[4]{2}$.

These results indicate, with good confidence, that the subset of random unitaries which was experimentally explored is a fair sample of the general random rotations set that may occur without a shared reference frame between $C$ and $B^C$.  Indeed, the numerical simulations show that the mean value expected for our experimental subset of unitaries is given by $\mean{\mathcal{B}_{subset}}\sim1.359$, quite close to the mean value of $\mean{\mathcal{B}}\sim 1.345$ obtained when exploring the whole set of unitary transformations  . Moreover both sets are bounded by the lower value of $\sqrt[4]{2}$, thus reinforcing that our experimental setup provides a good sampling of random unitaries.

\begin{figure*}[t]
\includegraphics[width=0.96\linewidth]{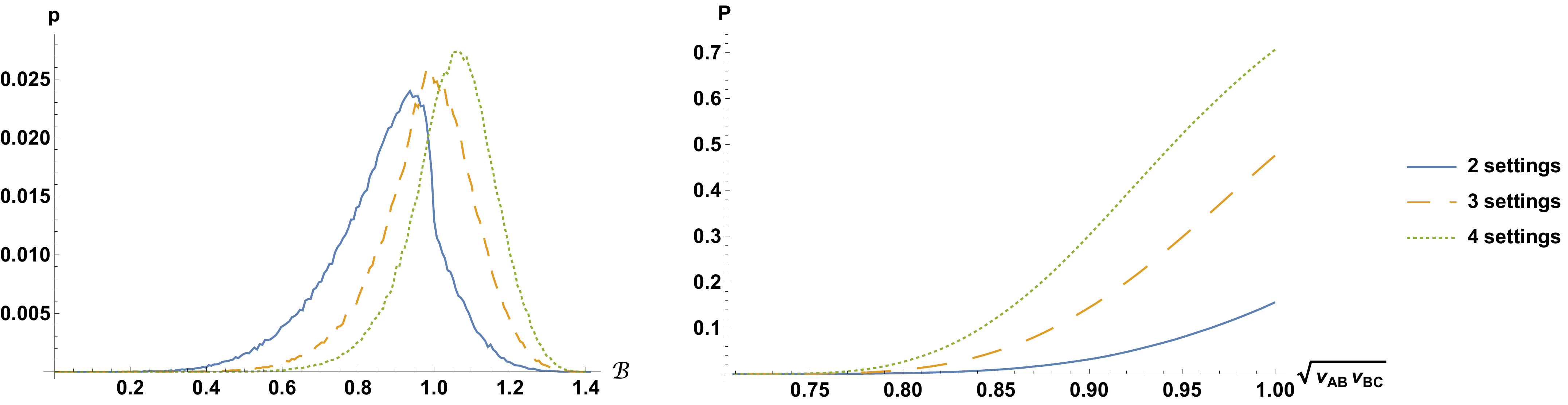}
\caption{\textbf{Numerical analysis of the case when no local calibration is ensured.} \textbf{a)} Probability density of obtaining a value $\mathcal{B}$. Different curves are related to an increasing number of settings (2,3,4) performed in stations A and C. \textbf{b)} Probability  violation with respect to visibility of the Werner states $\rho_{AB}$ and $\rho_{BC}$ for the same cases.}
\label{fig:No_Local_Calibration}
\end{figure*}

All these simulations were derived adopting a uniform distribution for the rotation parameters. Although this is not the most general distribution, we believe it is the most suitable to model the experimental case.

In \figref{fig:Main_experimental_plot} of the main text we presented our experimental data. It can be noticed that the orange triangle (corresponding to the identity transformation) exhibits a minimal violation of the bilocality inequality. In order to have a maximal violation (instead of a minimum point) in correspondence of the identity transformation, it is necessary to perform the measurements triad $\{\sigma_x,\;\sigma_y\;\sigma_z\}$ in station $B^C$ and $\{(\sigma_x+\sigma_z)/\sqrt{2},\;\sigma_y,\;(\sigma_x-\sigma_z)/\sqrt{2}\}$ in station C (or viceversa). However, we considered a case where there is no clue at all about the possible reference frames' shift, and all rotations have equal probability. In this situation, the most reasonable choice was performing the three orthogonal triads $\{\sigma_x,\;\sigma_y\;\sigma_z\}$ in both stations, also lowering the experimental demands.

\textit{No shared reference frames at all} - Numerical analysis of this case was also performed. This time we have $6$ independent variables associated to the relative shifts between $A,\;B^A$ and $C, \;B^C$, analogously to the previous case. We performed several independent simulations, testing an increasing population of $N$ random shifts. Our conclusions are shown in  \figref{fig:NoRefAtAll}. \figref{fig:NoRefAtAll}-a shows the clear convergence of the mean violation to the value of $\mean{\mathcal{B}^{3333}}\sim 1.297$, while \figref{fig:NoRefAtAll}-b gives reasonable evidence of the asymptotic lower bound $\min(\mathcal{B}^{3333})=1$. This asymptotic bound is, indeed, exactly what we expect considering that the inequality \ref{eq:No_ref_at_all_ineq} is not strict anymore in case all reference frames are aligned.

\textit{Absence of local calibration} - We numerically studied the case when neither shared reference frames nor local calibration are ensured. We analyzed the cases when stations $A$ and $C$ perform 2, 3 or 4 measurements, considering each time a population of $N=2\cdot 10^5$ runs. \figref{fig:No_Local_Calibration}-a shows the probability density associated to a value of $\mathcal{B}$. This function is obtained grouping the results of the $N$ runs in $200$ bins dividing the interval $0\leq \mathcal{B} \leq \sqrt{2}$. Curves with different colors are referred to an increasing number of measurements performed in stations $A$ and $C$. 

\figref{fig:No_Local_Calibration}-b shows instead the variation of the violation probability with respect to the noise parameter $V=\sqrt{v_{AB}v_{BC}}$. This figure is obtained integrating the probability densities of \figref{fig:No_Local_Calibration}-a from $1/V$ to $\sqrt{2}$. As we can see, increasing the number of measurements quickly makes the violation stronger to noise. This scheme, however, relies on the exploitation of an higher number of measurements in stations $A$ and $C$. It is reasonable, however, to guess that increasing also the measurements performed in substations $B^A$ and $B^C$, a stronger resistance of violation probability against noise can be obtained.

\bibliographystyle{apsrev4-1}
%\bibliography{bilocalexperiment}

\end{document}